\documentclass[12pt]{article}

\usepackage{amsmath}
\usepackage{amsfonts,amssymb,amsthm,overpic}
\makeatletter
\newcommand{\diam}{\mathop{\operator@font diam}}
\usepackage{setspace}
\usepackage{multicol}
\usepackage{multirow}
\usepackage[compact]{titlesec}

\usepackage{epsfig}

\newtheorem{remark}{Remark}[section]

\newtheorem{theorem}{Theorem}[section]

\newtheorem{corollary}{Corollary}[section]

\newcommand{\cT}{\mathcal{T}}

\begin{document}

\title{\Huge{\textsc{On two topologies that were suggested by Zeeman.}}}

\author{Kyriakos Papadopoulos$^1$, Basil K. Papadopoulos$^2$\\
\small{1. Department of Mathematics, Kuwait University, PO Box 5969, Safat 13060, Kuwait}\\
\small{2. Department of Civil Engineering, Democritus University of Thrace, Greece}}

\date{}

\maketitle

\begin{abstract}
The class of Zeeman topologies on spacetimes in the frame of relativity theory is considered to be of powerful intuitive justification, satisfying a sequence of properties with physical meaning, such as the group of homeomorphisms under such a topology is isomorphic to the Lorentz group and dilatations, in Minkowski spacetime, and to the group of homothetic symmetries in any curved spacetime. In this article we focus on two distinct topologies that were suggested by Zeeman as alternatives to his Fine topology, showing their connection with two orders: a timelike and a (non-causal) spacelike one. For the (non-causal) spacelike order, we introduce a partition of the null cone which gives the desired topology invariantly from the choice of the hyperplane of partition. In particular, we observe that these two orders induce topologies within the class of Zeeman topologies, while the two suggested topologies by Zeeman himself are intersection topologies of these two order topologies (respectively) with the manifold topology. We end up with a list of open questions and a discussion, comparing the topologies with bounded against those with unbounded open sets and their possible physical interpretation.
\end{abstract}

\section{Introductory Survey of the Terrain, with Comments.}

In this paper we focus entirely in the last section of Zeeman's paper \cite{Zeeman1}, and we discover the
link of two of the three topologies mentioned there with the Orderability Problem (see \cite{Good-Papadopoulos}). In particular, we
show that these two topologies are intersection topologies (in the sense of G.M. Reed, see \cite{Intersection}) of the manifold topology (topology of $\mathbb{R}^4$ in the case of Minkoski spacetime) with
(respectively) two order topologies (one induced by a chronological order and the other
by a non-causal, spacelike order). We have discussed about the remaining suggested topology by Zeeman in
the last paragraph of his paper
\cite{Zeeman1}, in articles \cite{OrderLightCone} and \cite{Ordr-Ambient-Boundary}.

\subsection{Causal Relations.}

In spacetime geometry it became standard to introduce three  relations, namely the
chronological irreflexive partial order $\ll$, the causal reflexive partial order $\prec$ and the reflexive relation horismos $\rightarrow$, and these can be extended to  any \emph{event-set}, that is a set
$(X,\ll,\prec,\rightarrow)$ equipped with all three of these relations having no metric (see \cite{Penrose-Kronheimer} and \cite{Penrose-difftopology}). We say that an event $x$ {\em chronologically precedes} event $y$, written
$x\ll y$ if $y$ lies inside the future null cone of $x$, $x$ {\em causally precedes} $y$,
$x \prec y$, if $y$ lies inside or on the future null cone of $x$ and  $x$ is at {\em horismos}
with $y$, written $x \rightarrow y$, if $y$ lies on the future null cone of $x$.  The notations $I^+(x) = \{y \in M : x \ll y\},  J^+(x) = \{y \in M : x \prec y\}$ will be used for the chronological and the causal futures  of $x$ respectively (and with a minus instead of a plus sign  for the pasts), while the future null cone of $x$ will be denoted by $\mathcal{N}^+(x)\equiv\partial J^{+}(x)= \{y \in M : x \rightarrow y\}$, and dually for the null past of $x$.

The above definitions of  futures and pasts of a set can be extended to the situation of any partially ordered set $(X,<)$.
In a purely topological context this is usually done by passing to the upper (we will call them future) and lower (we will call them past)
sets which in turn lead to the  future  and past  topologies (see \cite{Compendium} for the stronger case
of lattices; in our case, the standard definitions will apply in the particular case of $2$-dimensional
Minkowski spacetime, while in $4$-dimensional Minkowski or general relativistic spacetime manifolds our
topologies are weaker versions of the standard ones that appear in \cite{Compendium}).
A subset $A \subset X$ is a {\em past set} if $A = I^-(A)$ and dually for the future. Then, the {\em future topology} $\cT^+$ is generated
by the subbase $\mathcal{S}^+ = \{X \setminus I^-(x) : x \in X\}$
and the {\em past topology} $\cT^-$ by $\mathcal{S}^- = \{X \setminus I^+(x)  : x \in X\}$.
The (weak) {\em interval topology} $\cT_{in}$ on $X$ then consists of basic sets which are finite intersections
of subbasic sets of the past and the future topologies. This is in fact the topology that characterizes a given order of the poset $X$
and, in our case, the causal structure of a spacetime.

The \emph{orderability  problem} is concerned with the conditions under which the topology $\cT_<$ induced
by the order $<$  is equal to some given topology $T$ on $X$ (\cite{Good-Papadopoulos}, \cite{Orderability-Theorem},
\cite{OnProperties} and \cite{Nestsandtheirrole}).
In this paper, we present two special solutions to the orderability problem. In particular, we find that
two topologies suggested by Zeeman are actually intersection topologies (in the sense of G.M. Reed \cite{Intersection}) of
the manifold topology (topology of $\mathbb{R}^4$ in the case of Minkowski spacetime) with two order topologies (respectively) which belong again to the class
of Zeeman topologies, and we describe these order- (one of them non causal)
relations explicitly.

\subsection{The Zeeman Fine Topology.}

Zeeman (see \cite{Zeeman2} and \cite{Zeeman1}) showed that
the causal structure of the light cones on the Minkowski
spacetime determines its linear structure. After initiating
the question on whether a topology on Minkowski spacetime, which
depends on the null cones, implies its linear structure as well,
he constructed the Fine Topology (that we call here Zeeman $F$
topology) which is defined as the finest topology on $M$
which induces the $1$-dimensional Euclidean topology on every time
axis and the $3$-dimensional Euclidean topology on every space
axis.

Let $G$ be the group of automorphisms of the Minkowski space $M$, given by the Lorentz group,
translations and dilatations.

\begin{theorem}\label{main}
The group of homeomorphisms of the Minkowski space under $F$ is $G$.
\end{theorem}

Theorem \ref{main} gives to $F$ a great advantage, against the topology
of the Minkowski space whose group of homeomorphism is vast and of no
physical significance.

G\"{o}bel (see \cite{gobel}) showed that the results of Zeeman are valid
without any restriction on any spacetime, showing in particular that
the group of homeomorphisms of a spacetime $S$, with respect to the
general relativistic analogue of $F$, is the group of all homothetic
transformations of $S$.

Having G\"{o}bel's paper \cite{gobel} in mind, from now on we will denote
any spacetime with the letter $M$, without particularly restricting
ourselves to special relativity, unless otherwise stated.

In this paper we will focus on topologies which belong to the class of
Zeeman topologies but are coarser than $F$. Since we will also include
general relativistic analogues of Zeeman topologies, we will call
our topologies Zeeman-G\"obel topologies. In particular, a topology on a spacetime
is Zeeman-G\"obel if, when its open sets restricted to spacelike hyperplanes they give
three dimensional open sets while when restricted to timelike curves
they give one dimensional open sets, in the manifold topology.

\subsection{Light Cones, Time Cones and Space Cones.}

Let $M$ be the Minkowski space. If $x \in M$, the following
cones through $x$ are invariant under the group $G$ (Lorentz group and dilatations, \cite{Zeeman1}):

Light Cone: $C^L(x)= \{y: Q(y-x)=0\}$ \\
Time Cone: $C^T(x)=\{y: y=x \textrm{ or }Q(y-x)>0\}$\\
Space Cone: $C^S(x)=\{y: y=x \textrm{ or } Q(y-x)<0\}$,

where our metric signature is timelike, $(+,-,-,-)$, $Q$ denotes the characteristic quadratic form on $M$, given by
$Q = x_0^2-x_1^2-x_2^2-x_3^2$, $x= (x_0,x_1,x_2,x_3) \in M$
and $<$ is the partial order on $M$ given by $x<y$ if the
vector $y-x$ is timelike (this is actually the chronological order $\ll$ that we have mentioned previously).

In a curved spacetime $M$, under the frame of general relativity theory, we consider the existence
of a null cone for every event $x \in M$; so singular points, in this frame, do not belong
to our spacetime and are not of our interest. As soon as there exists spacetime manifold, there
are events. As soon as there is an event, there is a null cone; the partition of the null cone into
light cone (its topological boundary), time cone (its topological interior) and spacecone (its
topological exterior) helps to understand that the curvature of spacetime will {\em not} affect
the construction of our spacetime topologies. 

For simplicity, we will denote $C^L(x)$ by $L(x)$ (and the
corresponding null future and null past of $x$ by $L^+(x)$, $L^-(x)$, resp.) and
$C^T(x)$ by $T(x)$ (and $T^+(x)$, $T^-(x)$ the chronological
future and chronological past of $x$, resp.).

\subsection{A Partition of the Null Cone.}

For a two dimensional
spacetime $M$, we can divide the space cone $C^S(x)$ (denoted
by $S(x)$ for simplicity) into positive (right) space cone $S^+(x)$
and negative (left) space cone $S^-(x)$, where we do not particularly
need any consistency on what is ``right'' or ``left'', as
soon as the space cone is divided into two equal parts. For dimensions greater than $2$,
for such a division of $S(x)$ into $S^+(x)$ and $S^-(x)$ we need
to ``cut'' our null cones with a hyperplane passing trough
$x$ (in a vertical manner, dividing the causal cone of $x$ into
two equal parts ``left and right'' and so its space cone as well; for
an event $x$ and its causal cone $L(x) \cup T(x)$ one can choose
an appropriate event $y$ inside the cone, so that the line
$xy$ is contained in such a plane), and consider one ``slice'' of $S(x)$ as $S^+(x)$ and
the remaining one as $S^-(x)$. As we shall see, the division
of the space cone into two parts, in a symmetrical manner (i.e. two equal parts),
need not be done in a fixed and consistent way for every
event $x$; with respect to our study, we do not need to define a ``positive'' orientation
on the space cones in a dual way to time orientation. The idea of splitting space cones into ``right'' and ``left'' subcones
is important for our construction and study of a particular Zeeman topology,
which we shall show that it is an order topology, invariantly
of the way that one divides a space cone into ``left'' and ``right''
space cones (or ``positive'' and ``negative'' ones; in a $4$-dimensional
spacetime manifold there are infinitely many ways to divide a space
cone into two equal parts and we just need to fix one such way, arbitrarily.).

Here we should highlight what we mentioned in the previous subsection: since
our interest on the null cone is purely topological, as soon as there is a null
cone, one can partition it into interior, boundary and exterior, independently
on whether the null cone lies in a linear space or in a curved spacetime. Thus,
the ``cut'' of the null cone into ``two equal halves'' is not restricted in
the Minkowski space.

\section{On the Finest Topology that Induces the $1$-Dimensional Euclidean Topology on Every Time Axis.}

Having mentioned in Section 1.3 an idea to partition the space cone
$S(x)$ for an event $x$ in $M$ into two symmetrical subcones
$S^+(x)$ and $S^-(x)$, we will now introduce a space-like (non causal) order $<$,
such that $x <y$ if $y \in S^+(x)$ and $x>y$ if $y \in S^-(x)$.
By $x\leq y$ we mean that either $y \in S^+(x)$ or $x \rightarrow^{irr} y$
and, respectively, $y \leq x$ if either $y \in S^-(x)$ or $x \rightarrow^{irr} y$,
where $\rightarrow^{irr}$ denotes the irreflexive version
of horismos in $\leq$; so the order $\leq$ refers to space-like or light-like separation
of an event $x$
but excludes the case that $x \rightarrow x$.

We read from Zeeman (see the last section of \cite{Zeeman1}) that
the finest topology that induces the $1$-dimensional Euclidean topology on
every time axis is an alternative topology for the Zeeman fine topology, $F$.
We denote this topology by $Z^T$.
An open neighbourhood of $x \in M$ is given by the intersection of a Euclidean
open ball with center $x$ and the time cone of $x$, namely $B_{\epsilon}^E \cap C^T(x)$. It follows that
the induced topology on any space axis is discrete, which is suggestive of the discreteness of matter.
This topology  has a further physically attractive
feature as follows: If $f : I \to M$ is an embedding (not necessarily order preserving)
then $f(I)$ is a piecewise linear path along time axes, zig-zagging with respect to time
orientation, like the Feynman track of an electron. Theorem \ref{main} is also satisfied.

Here we observe that $Z^T$ is the intersection topology (in the sense of
G.M. Reed, \cite{Intersection}) of the interval topology $T_{in}^{\le}$
which is generated by the spacelike order $\le$ and the Euclidean topology $E^4$
on $\mathbb{R}^4$, in case of Minskowski spacetime, or of the manifold topology $M^4$ in any curved
spacetime. We note that, in the case of manifold topology, we define open balls (in a
similar way one defines Euclidean open balls) by some arbitrary Riemannian metric $h$,
on $M$.

In particular, from the definition of chronological order, we have that
$x \ll y$, if $x \prec y$ but not $x \rightarrow y$.
To see if $Z^T$ is the intersection topology of the mentioned interval topology and $M^4$ and, if so,
to find the order from which this interval topology is
induced, we first consider the complements of the sets:
\[ S^+(x) = \{y \in M : y > x \textrm{ or } y \rightarrow x\},~~~~~(1)\]
and dually for $M - S^-(x)$. Then, we observe that the intersection of the subbasic
sets $M - S^+(x)$ and $M - S^-(x)$ gives a $T_{in}^{\le}$ open set
and, consequently, a $Z^T$ open set if we intersect it with a ball $N^M_\epsilon(x)$ of $M^4$, because for each
$x$, $M - S^+(x) \cap M - S^-(x) \cap N^M_\epsilon(x)$ gives a neighbourhood $N^M_\epsilon(x)$ of some
radius $\epsilon$, with the space
and null cones removed, but $x$ is kept. We observe that
$T_{in}^{\le}$ also belongs to the class of Zeeman topologies. We thus, summarise as follows.

\begin{theorem}\label{1}
The order $\leq$ induces a Zeeman topology $T_{in}^{\le}$, which is
an interval topology generated by the spacelike order $\le$ and,
furthermore, $Z^T$ is the intersection topology of $T_{in}^\le$ and
$M^4$, on $M$. In addition, $T_{in}^{\le}$ (and, consequently, $Z^T$)
is defined invariantly of the choice of the partition of the null cone
into two, for defining $\le$.
\end{theorem}


\begin{remark}\label{remark important2}
Consider $x <y$, that is, $y \in S^+(x)$. What about if we partition $y$'s cone, so that
$y \in S^+(x)$? Since Theorem \ref{1} works invariantly from the way that cones are divided into two,
we can impose a condition, so that the spaceline order $\le$ is defined in our spacetime as
a strict order. In other words, if $y \in S^+(x)$, then $x \in S^-(y)$.
\end{remark}

\section{On the Finest Topology that Induces the $3$-Dimensional Euclidean Topology on Every Space Axis.}

According to Zeeman (see \cite{Zeeman1}), the finest topology that induces the $3$-dimensional Euclidean
topology on every space axis is another alternative topology for the Zeeman fine topology, $F$.
We denote this topology by $Z^S$. Two main properties of $Z^S$ are that the induced topology on every time axis is discrete
and that Theorem \ref{main} is satisfied.

Just as we did for $Z^T$, we show that $Z^S$ is the intersection
topology of $M^4$ with a Zeeman interval (order) topology
$T_{in}^{\ll^=}$, where $\ll^=$ is a shortcut for $\ll \cup \rightarrow$,
but $\rightarrow$ is considered irreflexive again, as in the previous paragraph.

We note that  $\ll^=$ is an irreflexive order, while
$\prec$ is reflexive.

From the definition of space cone, we have that
$y$ lies in the space cone of $x$, if $y$ neither lies
in the time cone of $x$ nor in the null cone of $x$.

To see if $Z^S$ is an intersection topology between an interval (order) topology and $M^4$ and if so,
to find the order from which this interval topology is
induced, we first consider the complements of the sets:
\[C^+(x) = \{y \in M : y \gg x \textrm{ or } y \rightarrow x\},~~~~~(2)\]

and dually for $M - C^-(x)$. Then, we observe that the intersection of the subbasic
sets $M - C^+(x)$ and $M - C^-(x)$ gives basic open sets in $T_{in}^{\ll^=}$,
the interval order topology induced by $\ll^=$. Then, intersecting
the basic open sets of $T_{in}^{\ll^=}$ with $M^4$ balls $N^M_\epsilon(x)$, we get $Z^S$ basic open sets, because for each
$x$, $M - C^+(x) \cap M - C^-(x)$ gives a neighbourhood in the manifold topology, with the time
and null cones removed, but $x$ kept. We have thus shown the following.

\begin{theorem}\label{2}
The order $\ll^=$ (that is, the irreflexive causal order) induces an interval
topology $T_{in}^{\ll^=}$ which belongs to the class of Zeeman topologies and the Zeeman topology $Z^S$ is the intersection
topology of $T_{in}^{\ll^=}$ and $M^4$, on $M$.
\end{theorem}

\section{On the Zeeman $Z$ Topology.}

We call $Z$ the topology that is mentioned by Zeeman
(see \cite{Zeeman1} and also \cite{OrderLightCone}) as an alternative topology for $F$. This topology is coarser than the Fine
Zeeman topology $F$, and it has a countable base of open sets of
the form:
\[ Z_\epsilon (x) = N_\epsilon^E(x) \cap (T(x) \cup S(x)) \]

The sets $Z(X)$ are open in $M^F$ (the manifold $M$ equipped with $F$) but not in
$E^4$ (Minkowski spacetime) or $M^4$. In addition,
the topology $Z$ is finer than $E^4$ or $M^4$ .  Theorem \ref{main} is satisfied, among
other properties that $F$ satisfies as well. According to Zeeman, $Z$ is technically simpler than $F$,
but it is intuitively less attractive than $F$.



Theorems \ref{1} and \ref{2} give rise to the following corollary.

\begin{corollary}\label{corollary 1}

Consider the topologies $Z^T$, $Z^S$ and $Z$. Consider basic-open
sets $Z^T(x) \in Z^T$, $Z^S(x) \in Z^S$ and $Z(x) \in Z$. Then:
\begin{enumerate}

\item $Z^T(x) \cup Z^S(x) = Z(x)$ and

\item the intersection of the order relations that induce the interval topologies $T_{in}^\le$ and $T_{in}^{\ll^=}$,
namely $\leq$ intersected with $\ll^=$, equals the irreflexive horismos $\rightarrow^{irr}$.

\end{enumerate}

\end{corollary}

\section{Discussion and Open Questions.}

In our conclusions succeeding Theorem 2.1
in paper \cite{Ordr-Ambient-Boundary}, we stated that
``Neither the chronological order $\ll$ nor the
causal order $\prec$ induce a topology equal to the Zeeman
topology'' (in this case we meant the Zeeman $Z$ topology).
The rest of this mentioned paragraph refers to the existence of
interval topologies generated by $\ll$ and the irreflexive $\prec$ (denoted by $\ll^= here)$,
respectively, ignoring that they are Zeeman topologies
as well (see section 7, of article \cite{Zeeman1}). Theorem \ref{1} and Theorem \ref{2}
give us that both topologies are intersection topologies of interval topologies with $M^4$ as well, as speculated
in \cite{Ordr-Ambient-Boundary} and, so, they are connected with the
order relations that were mentioned, namely the spacelike one $\le$ and the irreflexive causal $\ll^=$.

In an environment under $Z^T$, the dominant order (that is, the one which
is related to the topology) is $\leq$.
This is not a causal order and, although time continues to be one of the
coordinates, it loses any relation to the idea of causality as it is known
in general relativity. Only null or spacelike relations are permitted. In a sense,
time is not felt by an observer, while massless particles like photons and gravitons
move on the null cone, or events are joint with spacelike curves.

{\bf{Question 1:}}
As the division of the spacecone $S(x)$ ``into two equal subcones'' is done with no restriction (the interval topology we get will be
the same, no matter which of the appropriate hyperplanes through $x$ we choose to divide the spacecone), it is not clear yet how
the spacelike relation $\leq$ affects the chronological order $\ll$ of the
time cone $T(x)$, but it does affect it, indeed. We emphasize that the order $\leq$ is not meant to
be causal (since otherwise it would refer to objects with speed greater than the speed of light),
but it is an order relation related to the matter. The order-theoretic properties
of $\leq$, and its physical meaning under the frame of general relativity,
should be studied explicitely, since $\leq$ is the order of the interval
topology $T_{in}^{\leq}$ which is Zeeman itself and it is closely related to $Z^T$. In addition, if we wish $\le$ to be a strict order, we can
add the restriction given in Remark \ref{remark important2} which, again, does
not affect the topology of the space. This raises an even more difficult
question: $\le$ and $\le$ under Remark \ref{remark important2} certainly
affect space in a different way, but they both influence time in the same
way. Why is this, at least from a topological perspective?

In an environment under $Z^S$, the order that induces the topology is the irreflexive causal order $\prec$,
that we denoted by $\ll^=$.
Here there is no violation in the causal structure
of $M$ as is known in general relativity.

{\bf{Question 2:}} We now ask the
reverse of Question 1. How is the causal order $\ll^=$ (related to $Z^S)$ affecting the
spacelike order $\leq$ of the space cone of $x$.

In an environment under $Z$, the dominant relation related to the topology is $\rightarrow^{irr}$.
As we discussed in \cite{Ordr-Ambient-Boundary}, causality is different from as it is known in general relativity. In a sense, time
is not felt by any observer, while massless particles move on the null cone.

{\bf{Question 3:}} Corollary \ref{corollary 1} gives rise to questions of a bit deeper
nature. Properties 1. and 2. make us wonder on how a combination of time- and
space- open sets give rise to an order referring to light cones, i.e. to massless
objects. Here we have a given topological condition affecting in a certain
way the causality; the whole setting seems to be unexplored in terms of a physical interpretation.
It is worth mentioning (as was also mentioned in \cite{Ordr-Ambient-Boundary}) that
in an environment under $\rightarrow$ (or $\rightarrow^{irr}$), the causal curves are piece-wise null curves.

{\bf{Question 4:}} It would be interesting to see under which spacetimes
will the topologies $Z^T$, $Z^S$ and $Z$ play a significant role. It seems
that $Z$ is meaningful in Planck time environment and creates an ``ideal''
environment for the study of achronal sets.

{\bf{Question 5:}} A topological space itself can be considered as a ``static''
mathematical object. It would be more realistic, in the frame of spacetime
in general relativity, if we consider a dynamical evolution of Zeeman topologies,
in a global and local manner. $Z$ could be a candidate for environments
``close to'' a blackhole, for example, while $F$ is the largest
topology in a spacetime, satisfying the maximum number of physically attractive properties.

\end{document}